\documentclass[twocolumn,aps,superscriptaddress]{revtex4-1}
\usepackage{palatino}
\usepackage[latin1]{inputenc}
\usepackage{epsf}
\usepackage{amsmath,amssymb}
\usepackage{latexsym}
\usepackage{calc}
\usepackage{color}
\usepackage{shadow}
\usepackage{epsfig}

\newcommand{\ben}{\begin{equation}}
\newcommand{\een}{\end{equation}}
\newcommand{\bea}{\begin{eqnarray}}
\newcommand{\eea}{\end{eqnarray}}

\def\dulR{{\underline{\underline{\bf R}}}}
\def\dulr{{\underline{\underline{\bf r}}}}

\def\dulS{{\underline{\underline{\sigma}}}}
\def\duls{{\underline{\underline{\bf s}}}}

\def\na{{\nabla}}

\begin{document}
\title{Reply to ``Comment on 'Correlated electron-nuclear dynamics: Exact factorization of the molecular wavefunction' [J. Chem. Phys. {\bf 137}, 22A530 (2012) ]''}
\author{Ali Abedi}
\affiliation{Max-Planck Institut f\"ur Mikrostrukturphysik, Weinberg 2, D-06120 Halle, Germany}
\affiliation{European Theoretical Spectroscopy Facility (ETSF)}
\author{Neepa T. Maitra} 
\affiliation{Department of Physics and Astronomy, Hunter College and the City University of New York, 695 Park Avenue, New York, New York 10065, USA}
\author{E.K.U. Gross}
\affiliation{Max-Planck Institut f\"ur Mikrostrukturphysik, Weinberg 2, D-06120 Halle, Germany}
\affiliation{European Theoretical Spectroscopy Facility (ETSF)}
\date{\today}
\maketitle

Ref.~\cite{ACEJ13} raises the question of whether the equations for
the electronic and nuclear wavefunctions derived in Ref.~\cite{AMG12}
preserve the norms of these wavefunctions, and provides a proof that
the equations do in fact preserve the norms. There is another way to
see that the norms are preserved by the time evolution, which was
indicated in the original earlier work Ref.~\cite{AMG10}, although it
was not fleshed out in Ref.~\cite{AMG12}.  In this earlier work where
the exact factorization was first presented, the derivation of the
equations is outlined in the following statement: "We require the
action to be stationary with respect to variations in
$\Phi_\dulR(\dulr,t)$ and $\chi(\dulR,t)$ subject to the condition
(5)." where condition (5) is the partial normalization condition (PNC) of
Ref.~\cite{AMG10}, and the action is the Frenkel action ${\cal
  S}[\Psi,\Psi^*] = \int_{t_i}^{t_f} dt\langle\Psi\vert \hat{H} -
i\partial_t\vert \Psi\rangle$, with $\hat{H}$ the Hamiltonian of the
full molecular system.  A common practise in variational methods is to
enforce such conditions via the method of Lagrange multipliers.
Ref.~\cite{AMG12} provides the details of the stationarizing
procedure, but the derivation appears to neglect the term that
enforces the normalization constraint. In fact, the Lagrange
multiplier turns out to be zero as we shall show below explicitly.

One adds a term with Lagrange
multiplier $\mu(\dulR\dulS,t)$ to the action of Eq. (37) in Ref.~\cite{AMG12}, to enforce the PNC, defining
\bea
\nonumber
{\tilde{\cal S}}[\Phi_{\dulR\dulS},\Phi_{\dulR\dulS}^*, \chi, \chi^*] = {\cal S}[\Phi_{\dulR\dulS},\Phi_{\dulR\dulS}^*, \chi, \chi^*]\\
+\sum_\dulS\int d\dulR  \mu(\dulR~\dulS,t)\sum_\duls\int d\dulr |\Phi_{\dulR\dulS}(\dulr\duls,t)|^2,
\eea
and performs the variations of ${\tilde{\cal S}}[\Phi_{\dulR\dulS},\Phi_{\dulR\dulS}^*, \chi, \chi^*]$ with respect to the electronic and nuclear wavefunctions. 
Requiring $ 0= \delta\tilde{\cal S}/\delta \chi^*$ yields
%\ben
%\langle\Phi_{\dulR\dulS}\vert \hat{H}_{BO} + \hat{V}^e --- \vert\Phi_{\dulR\dulS}\rangle_{\dulr\duls}\chi + \langle\Phi_{\dulR\dulS}\vert\Phi_{\dulR\dulS}\rangle_{\dulr\duls}\left(-----\right)\chi + \frac{1}{M}....
%eqn 1 of mu1
\begin{widetext}
\ben
\label{eq:variationchi}
\left\langle \Phi_{\dulR~\dulS}\right\vert \hat{H}^{new}_e-i\partial_t\left\vert\Phi_{\dulR~\dulS}
\right\rangle
 \chi + \left[\sum_{\alpha}\frac{1}{M_{\alpha}} (-i\nabla_\alpha\chi/\chi)
\cdot{\bf A}_{\alpha}\right] \chi = -\langle\Phi_{\dulR\dulS}\vert\Phi_{\dulR\dulS}\rangle\left[\sum_{\alpha} \frac{-\na_{\alpha}^2}{2M_{\alpha}}+\hat{V}_{ext}^n-i \partial_t\right] \chi
\een
while requiring
$ 0= \delta\tilde{\cal S}/\delta \Phi_{\dulR\dulS}^*$ yields 
\ben
\label{eq:variationphi2}
\left(\hat{H}^{new}_e-i\partial_t +\mu/\vert\chi\vert^2\right)\Phi_{\dulR~\dulS} + \sum_{\alpha} \frac{1}{M_{\alpha}}(-i\nabla_\alpha\chi/\chi)\cdot (-i\na_{\alpha} \Phi_{\dulR~\dulS})
= - \frac{(\sum_{\alpha} \frac{-\na_{\alpha}^2}{2M_{\alpha}} + \hat{V}_{ext}^n-i\partial_t) \chi} {\chi} \cdot \Phi_{\dulR~\dulS}
\een
where $\hat{H}^{new}_e \equiv \hat{H}_{BO} +\hat{V}_{ext}^e+ \sum_{\alpha} \frac{-\na_{\alpha}^2}{2M_{\alpha}}$ and the notation follows that of 
Ref.~\cite{AMG12}. Replacing the right-hand-side of Eq.~(\ref{eq:variationphi2}) by the LHS of Eq.~(\ref{eq:variationchi}), multiplied by
$\Phi_{\dulR\dulS}/\left(\langle\Phi_{\dulR\dulS}\vert\Phi_{\dulR\dulS}\rangle\cdot\chi\right)$ we arrive at:
\ben
\label{eq:elec}
\left[\left(\hat{H}^{new}_e-i\partial_t +\frac{\mu}{\vert\chi\vert^2}\right)\Phi_{\dulR~\dulS} + \sum_{\alpha} \frac{1}{M_{\alpha}}(-i\nabla_\alpha\chi/\chi)\cdot (-i\na_{\alpha} \Phi_{\dulR~\dulS})\right]
= \frac{\left[\left\langle \Phi_{\dulR~\dulS}\right\vert \hat{H}^{new}_e-i\partial_t\left\vert\Phi_{\dulR~\dulS}\right\rangle + \sum_{\alpha}\frac{1}{M_{\alpha}} (-i\nabla_\alpha\chi/\chi)
\cdot{\bf A}_{\alpha}\right]\Phi_{\dulR~\dulS}}{\langle\Phi_{\dulR\dulS}\vert\Phi_{\dulR\dulS}\rangle}
\een

\end{widetext}

At first sight the procedure appears to add a term $\mu(\dulR\dulS,t)$ to the resulting electronic
Hamiltonian.
However, multiplying Eq.~(\ref{eq:elec}) by $\Phi_{\dulR\dulS}^*(\dulr\duls,t)$ and integrating over $\dulr\duls$, readily determines the value of the Lagrange
multiplier: $\mu(\dulR\dulS,t) = 0$. 

Eq.~(\ref{eq:elec}) was derived 
using the method of Lagrange Multipliers to enforce the PNC: it
therefore preserves the norm, and so starting with a state
$\Phi_{\dulR\dulS}(0)$ that has unit norm
$\langle\Phi_{\dulR\dulS}(0)\vert\Phi_{\dulR\dulS}(0)\rangle_{\dulr\duls}
=1$ means that it will evolve with unit norm
$\langle\Phi_{\dulR\dulS}(t)\vert\Phi_{\dulR\dulS}(t)\rangle_{\dulr\duls}
=1$. Since $\mu(\dulR\dulS,t) = 0$, Eq.~\ref{eq:elec} then reduces to
the electronic equation (28) of Ref.~\cite{AMG12}. There is also no
change in the nuclear equation. Hence, performing the
variation of the action under the partial normalization constraint
yields no change to any of the equations or statements in
Ref.~\cite{AMG12}. Including or not including the partial normalization 
condition as Lagrangian constraint in the variation of the action functional 
makes no difference in the resulting equations of motion because the Lagrangian 
multiplier $\mu(\dulR\dulS,t)$ vanishes identically.

{\it Acknowledgements} Partial support from the Deutsche
Forschungsgemeinschaft (SFB 762), the European Commission
(FP7-NMP-CRONOS), and from the National Science Foundation
(CHE-1152784) (NTM) is gratefully acknowledged.

\end{document}